\documentclass[twocolumn,showpacs,prb]{revtex4}
\usepackage{graphicx}%
\usepackage{dcolumn}
\usepackage{amsmath}
\usepackage{latexsym}
\usepackage{longtable}
\begin{document}
\title{Crystal structure and physical properties of half-doped manganite nanocrystals with  size $<$ 100nm}
    
\author{Tapati Sarkar\footnote[1]{email:tapatis@bose.res.in}, Barnali Ghosh and  A. K. Raychaudhuri\footnote[2]{email:arup@bose res.in}}
\affiliation{DST Unit for NanoSciences, S.N.Bose National Centre for Basic Sciences, Block JD, Sector III, Salt Lake, Kolkata 700 098, West Bengal, India.}
\author{Tapan Chatterji}
\affiliation{Science Division, Institut Laue-Langevin, BP 156, 38042, Grenoble, Cedex 9, France.}

\date{\today}
\begin{abstract}

\noindent
In this paper we report the structural and property (magnetic and electrical transport) measurements of nanocrystals of half-doped $\mathrm{La_{0.5}Ca_{0.5}MnO_3}$(LCMO) synthesized by chemical route, having particle size down to an average diameter of $15$nm. It was observed that the size reduction leads to change in crystal structure and the room temperature structure is arrested so that the structure does not evolve on cooling unlike bulk samples. The structural change mainly affects the orthorhombic distortion of the lattice. By making comparison with observed crystal structure data under hydrostatic pressure it is suggested that the change in the crystal structure of the nanocrystals occurs due to an effective hydrostatic pressure created by the surface pressure on size reduction. This not only changes the structure but also causes the room temperature structure to freeze-in. The size reduction also does not allow the long supercell modulation needed for the Charge Ordering, characteristic of this half-doped manganite, to set-in. The magnetic and transport measurements also show that the Charge Ordering (CO) does not occur when the size is reduced below a critical size. Instead, the nanocrystals  show  ferromagnetic ordering down to the lowest temperatures along with metallic type conductivity. Our investigation establishes a structural basis for the destabilization of CO state observed in half-doped manganite nanocrystals.

\end{abstract}
\maketitle

\noindent
\section{\bf INTRODUCTION}
The doped perovskite oxide manganites (with $ABO_3$ structure) are fascinating because they can readily be tuned between  different electronic phases by proper substitution of cations. The ground state of the manganites can have distinct phases (a ferromagnetic(FM)  metal, a charge - ordered (CO) insulator or a paramagnetic polaron liquid) which are energetically close\cite{ref1}. The manganites contain interactions of different types that are often of comparable strengths\cite{ref2}. As a result the ground state of the manganites can be of different types depending on which of the interactions win over. The tuning of the ground state can be done by such factors as the carrier density, magnetic field, pressure and bi-axial strain. Of particular importance for the present investigation is the charge and orbitally ordered state observed in half-doped manganites that have equal amounts of Mn$^{3+}$ and Mn$^{4+}$  ions. The half-doped manganites give rise to charge  ordered state  associated with a real space ordering of 1:1 Mn$^{3+}$/Mn$^{4+}$ species accompanying a  structural change\cite{ref3}. Charge and orbitally ordered insulating state (COI) occurs in many half doped manganites like $\mathrm{La_{0.5}Ca_{0.5}MnO_3}$\cite{ref4}, $\mathrm{Pr_{0.5}Ca_{0.5}MnO_3}$\cite{ref5}, $\mathrm{Pr_{0.5}Sr_{0.5}MnO_3}$\cite{ref6}, $\mathrm{Nd_{0.5}Ca_{0.5}MnO_3}$\cite{ref7}, $\mathrm{Nd_{0.5}Sr_{0.5}MnO_3}$\cite{ref8}. 

\noindent
In this paper we investigate systematically the crystal structure and physical properties of nanocrystals of a specific half-doped manganite, $\mathrm{La_{0.5}Ca_{0.5}MnO_3}$, with size down to $15nm$ and study how  size reduction to the nanometer dimensions  can lead to a tuning of the ground state in these materials. $\mathrm{La_{0.5}Ca_{0.5}MnO_3}$ has a transition from paramagnetic to ferromagnetic state around 225K, followed by the charge ordering transition at $T= 155K$\cite{Mathur}. An antiferromagnetic order accompanies the charge ordering transition. In particular, we study the crystal structure of these nanocrystals using Synchrotron X-ray to establish the role of structure in tuning the physical properties. In perovskite oxides in general a specific ordered state is generally favored by a specific crystal structure. Thus, in these nanocrystals, the study of the role of crystal structure in stabilizing a specific ground state is of utmost significance.  

\noindent
The insulating CO state can be destabilized to a ferromagnetic metallic (FM) phase by a number of external perturbations that include magnetic field\cite{ref12}, doping, biaxial strain and pressure\cite{ref13} and in some cases even electric field\cite{ref14,ref15}. In this work, we have investigated what happens to the structure and physical properties of $\mathrm{La_{0.5}Ca_{0.5}MnO_3}$ when we bring down the sizes from bulk ($\sim$ 3.6$\mu$m) to sizes as small as 15$nm$, a change in size of more than two orders of magnitude. Earlier reports of studies on half doped manganites were done with particle sizes which were much larger and in some cases, in excess of $100 nm$\cite{ref16,bhat,das}. These studies have reported modification of the charge ordered state in these systems. However, no analysis of the crystal structure using high resolution X-ray diffraction data (as has been done here) has been made. We show that analysis of the crystal structure using high resolution powder diffraction data, helps us to understand some of the crucial factors that lead to destabilization of the COI state in these manganites below a certain critical size. In particular, we show that an equivalent hydrostatic pressure (arising from the surface pressure) can lock the room temperature structure and not allow it to evolve on cooling as is required for the CO to set-in. To our knowledge, this is the first study of the structural evolution of nanoparticles of manganites using high resolution diffraction techniques. A preliminary report of the structural data has been made recently by us\cite{APL}.

\noindent
\section{EXPERIMENTAL}
We have adopted a polymeric (polyol) precursor route to synthesize $\mathrm{La_{0.5}Ca_{0.5}MnO_3}$ (LCMO) nanocrystals with sizes down to 15nm. This method allows synthesis at a significantly lower sintering temperature compared to the conventional solid state procedure. In this technique the polymer (ethylene glycol in our case) helps in forming a close network of cations from the precursor solution and assists the reaction, enabling phase formation at relatively low temperatures\cite{ref17}. In a typical synthesis process, high purity ($>$99{\%})(CH$_{3}$COO)$_{3}$La$\cdot$1.5H$_{2}$O, Ca(CH$_{3}$CO$_{2})_{2}\cdot$H$_{2}$O and (CH$_{3}$COO)$_{2}$Mn$\cdot$4H$_{2}$O (procured from Sigma Aldrich\cite{ref18}) were dissolved in the desired stoichiometric proportions in acetic acid and water. To this solution an appropriate amount of ethylene glycol (molecular weight = 62.07 gm/mol) was added and heated till the sol was formed. The gel was dried overnight at $\approx$150\r{ }C. Pyrolysis was done at  350\r{ }C and 450\r{ }C followed by a  sintering at $\approx$650\r{ }C to obtain the desired chemical phase. The water - ethylene glycol ratio, heat - treatment employed during gelling, pyrolization and calcination were found to influence the particle size of the final product. We optimized these process parameters to obtain phase pure LCMO nanocrystals with a particle size of $\sim$15 nm (as established from XRD results, using Williamson Hall plot\cite{ref19} and Transmission Electron Microscopy (TEM) images). If the experiments can be carried out without pellet formation like the Synchroton X-Ray studies, the nanocrystals with $\sim$15$nm$ diameter can be used. We used pellets of the nanocrystals both for transport and magnetic measurements. These nanocrystals have been used for making samples of larger size by heat treatment. The  pellets  were sintered  at  different  temperatures  varying  from  650\r{ }C  to  1300\r{ }C  and  for  varying  time  periods  (5 hrs - 30 hrs). The sintering at different temperatures and times lead to particle growths to different sizes making it possible to grow grain sizes as large as 3.6 $\mu$m or more. This ensures that the particles with different sizes have the same chemical stoichiometry since they have been prepared from the same batch of starting nanocrystals. All  the  synthesized  samples  were  characterized  using  powder  X-Ray  Diffraction (XRD) using CuK$\alpha$  radiation  at  room temperature to estbalish the purity of the chemical phases. The  stoichiometry  of  the  nanopowder  was  also checked independently by a quantitative analysis  using  Inductively  Coupled  Plasma  Atomic  Emission  Spectroscopy  (ICPAES). The  pellets  were  also checked  for  oxygen  stoichiometry  using  iodometric  titration. Microstructural  characterization of the pellets  was  done  using  Field Emission Gun - Scanning  Electron  Microscopy  (FEG-SEM) including Energy dispersive X Ray analysis (EDAX)  and Transmission Electron Microscopy (TEM).   

\noindent
The  magnetic  measurements have  been  carried  out using a Quantum Design SQUID magnetometer\cite{ref20} and also a home made low field a.c. susceptibility bridge working at 33.33Hz. The  resistivity  measurements  were  done  using  standard  d.c.  four-probe  technique  in  the  temperature  range  4.2 K to 300 K in a closed cycle refrigerator\cite{ref21}. Magnetoresistance measurements were carried out in the temperature range 4.2 K to 300 K under a field of 10 T using a bath type cryostat.

\noindent
High resolution powder diffraction data were obtained at the European Synchrotron Radiation Facility, Grenoble, France using the BM-01B beamline and a wavelength of 0.375{\AA}, over the temperature range 5K-300K. The samples were in powder form and were kept in a borosilicate capillary. The calibration was done using standard Si samples. To extract the lattice parameters and study the structural evolution of the samples as a function of temperature, we used   nearly 5000-6000 points/scan using 6 detectors. The Rietveld analyis of the lattice structure were done using the FullProf Suite software\cite{FP}.

\section{RESULTS}
\subsection {CHARACTERIZATION-STRUCTURE AND STOICHIOMETRY}

We synthesized 6 different samples having particle size ranging from $15nm$ to $3660nm$ by pelletizing the as prepared powder and subjecting them to different annealing conditions as described before (see Table~\ref{tab:table1}). The  XRD  data  were  used  to  check that the samples prepared were of pure phase and to estimate  the  average  particle  size for the samples with lower particle size (see Table~\ref{tab:table1}). For larger particle size samples, the sizes  were estimated from FEG-SEM  images. The resolution of the FEG-SEM is 2 nm at the beam voltage used (5 kV). In the region of overlap, there is good agreement between the average size determined by the FEG-SEM and the XRD. Table \ref{tab:table1} also shows the unit cell volumes of the nanocrystals as obtained from the X-Ray data taken at room temperature. (Analysis of the powder diffraction data are given in a separate sub-section below).

\begin{table*}
\caption{\label{tab:table1}Annealing conditions, particle size and cell volume of the LCMO samples}
\begin{ruledtabular}
\begin{tabular}{ccccc}
Sample ID& Annealing temperature (\r{ }C)& Annealing time (hrs)& Particle size (r) (nm)&Cell Volume ({\AA}$^{3}$)\\
\hline
A&650&5&15&221.0\\
B&750&10&43&221.1\\
C&850&10&141&222.5\\
D&950&10&378&224.6\\
E&1100&30&600&224.6\\
F&1300&2&3660&224.7\\
\end{tabular}
\end{ruledtabular}
\end{table*}

\begin{figure}[t]
\begin{center}
\includegraphics[width=8cm,height=7cm]{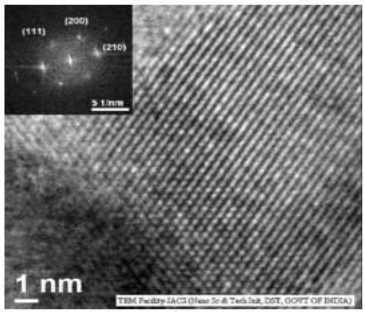}
\end{center}
\caption{HRTEM data of LCMO crystal of size 15nm. The lattice parameter obtained is $\sim$0.345 nm. The inset shows a typical diffraction pattern from a nanocrystal. The indexing has been done using orthorhombic symmetry.}
\label{Fig1}
\end{figure}

\noindent
In  Fig.~\ref{Fig1}  we  show  the  high  resolution  TEM  (HRTEM)  images taken on nanocrystals of size $\sim$ 15 nm.  The inset shows the electron diffraction pattern, which confirms the single crystalline nature of the nanoparticles. The indexing of the points in the diffraction pattern was done following orthorhombic symmetry. The lattice planes have been indexed. The   value  of  d ((111) planes)  as  obtained  from  the  HRTEM  images  is  $\sim$0.345 nm. This is quite close to the value obtained from the XRD ($\sim$0.343 nm) data.

\noindent
The  charge  ordering  phenomenon  observed in $\mathrm{La_{1-x}Ca_{x}MnO_3}$\cite{ref1} occurs  at  half doping with the carrier concentration $x \approx 0.5$.  We have taken special care that the samples used have proper stoichiometry and hence the proper carrier concentration. Since we made all the samples starting from the same batch of nanoparticles (with size $\sim$ 15nm), they all have the same stoichiometry and the same carrier concentration irrespective of the size.  We  checked  the stoichiometry  using  Inductively  Coupled  Plasma  Atomic  Emission  Spectroscopy  (ICPAES). The particular batch of samples on which we report the data here yielded  a  La:Ca:Mn  ratio  of  0.507:0.495:1.  We have checked that our synthesis  method  maintains  a  reproducible and good  stoichiometry from batch to batch. The control of proper stoichiometry among the samples with different sizes allows us to be   conclusive   that  any  change of physical  property  arises  only  because  of  a  reduction  in the size. In addition, we have also checked the stoichiometry on the microstructural level from EDAX measurement done on each pellet at different points over a range of $\sim$ 1 $\mu$m using the FEG-SEM. The spectral resolution was $\approx 10nm$. We find that the composition remained the same (the ratio La:Ca varies between 1.02 $\pm$ 0.04 throughout the scanned range in the samples used). We  also checked  the  oxygen  stoichiometry  of  the  pellets  by  iodometric  titration.  All the  pellets  had some  oxygen  deficiency  ($\mathrm{La_{0.5}Ca_{0.5}MnO_{3-\epsilon}}$)  with $\epsilon$ positive. $\epsilon \approx 0.021$ for the pellet with the smallest particle size  and it increases somewhat  for the bulk sample. Thus the particles with smaller particle size have better oxygen stoichiometry.

\noindent
\subsection{CRYSTAL STRUCTURE ANALYSIS}

\begin{figure}[t]
\begin{center}
\includegraphics[width=8cm,height=7cm]{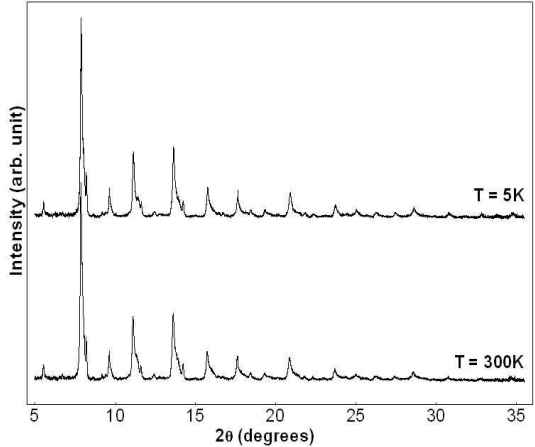}
\end{center}
\caption{Diffraction data of the nanoparticle sample (average particle size $\sim$ 15nm) taken at 5K and 300K.}
\label{Fig2}
\end{figure}

\begin{figure}[t]
\begin{center}
\includegraphics[width=8cm,height=7cm]{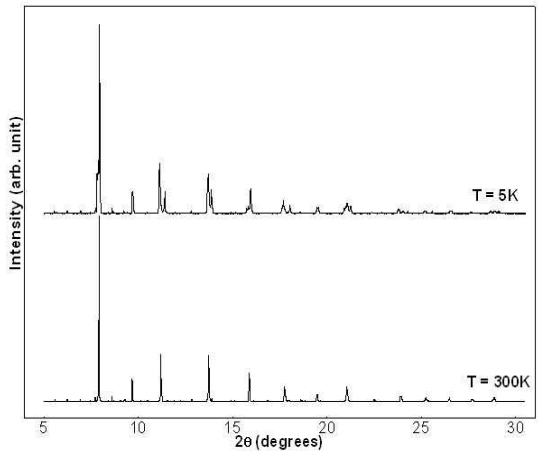}
\end{center}
\caption{Diffraction data of the bulk sample (average particle size $\sim 3.6\mu m$) taken at 5K and 300K.}
\label{Fig3}
\end{figure}

\noindent
In Fig.~\ref{Fig2} and Fig.~\ref{Fig3} we show the diffraction data taken at 5K and 300K for the nanocrystal (average diameter $\approx 15nm$) and the "bulk" (powder with grain size$\approx 3.6 \mu m$) respectively. For the bulk sample, the data taken at 300K is very different from that taken at the lowest most temperature of 5K. Here, we very clearly see the appearance of extra weak reflections in the x-ray diffraction patterns which are the signatures of small structural distortions occurring due to charge and orbital ordering. The appearance of these extra Bragg reflections have been associated with the presence of J-T distortion of the $Mn^{+3}O_6$ octahedra\cite{ref4}. On the other hand, the diffraction pattern of the nanocrystal remains virtually unchanged throughout the temperature region scanned i.e. the scan taken at 300K is virtually the same as that taken at 5K with no appearance of any extra peaks.  This itself is the first indication that the structure of the nanocrystals fails to evolve as it should on lowering the temperature if charge and orbital ordering has to set in. 

\begin{figure}[t]
\begin{center}
\includegraphics[width=8cm,height=7cm]{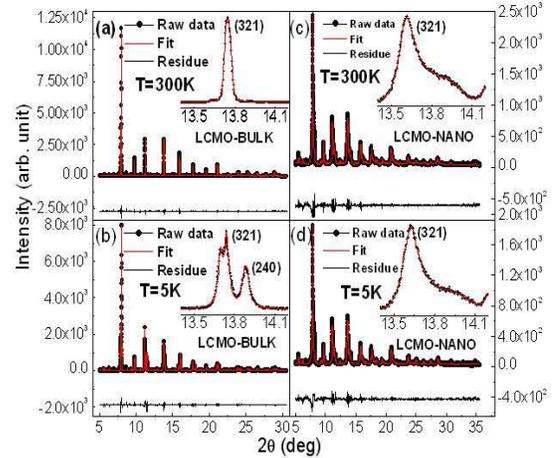}
\end{center}
\caption{XRD patterns (alongwith the fits) of (a)LCMO-BULK (at T = 300K), (b)LCMO-BULK (at T = 5K), (c)LCMO-NANO (at T = 300K) and (d)LCMO-NANO (at T = 5K). Insets show the expanded regions between $2\theta$ = 13.4\r{ } and 14.2\r{ }.}
\label{Fig4}
\end{figure}

\noindent
The high resolution diffraction data of the samples were analyzed using a profile fitting technique to obtain the crystal structure parameters. The X-ray line profiles were modeled using a Pseudo-Voigt profile shape function.  We show some representative fits along with the residues for both the samples in Fig.~\ref{Fig4}. This analysis was done to quantatitively study the  changes in the basic structure of $\mathrm{La_{0.5}Ca_{0.5}MnO_3}$ as a result of the size reduction. One obvious difference arising due to size reduction is in the shape of the peaks, the peaks in the nanoparticle sample being more broad and asymmetric than those in the bulk sample. This difference in the line shape can be taken care of through the shape parameter $\beta$. For the bulk sample, $\beta_{bulk}$ = 0.08 and that for the nanocrystals, $\beta_{nano}$ = 0.61 at T = 300K. On cooling, $\beta$ increases for both samples ($\beta_{bulk}$ = 0.21 and $\beta_{nano}$ = 0.66 at T = 5K), but the increase is much more in the bulk sample than that in the nanoparticle sample. Here we note that we did the analysis using the higher symmetry space group $Pnma$ at all temperatures.Below the charge ordering temperature the structure of the bulk sample undergoes an orthorhombic to monoclinic transition which will reduce the crystallographic symmetry from $Pnma$ to $P2_1/m$\cite{ref4}. However, here we report the refined values of the lattice parameters only (and not the atomic positions), and so the reported values are quite reliable. In fact, fitting the low temperature data using the space group $P2_1/m$ changes the refined values of the lattice parameters by only $\sim$ 0.0016\% (these values are not reported here). This change is lower than the \% error of 0.03\%. So effectively the values of the lattice parameters become independent of whether we use the space group $Pnma$ or $P2_1/m$. 

\begin{figure}[t]
\begin{center}
\includegraphics[width=8cm,height=7cm]{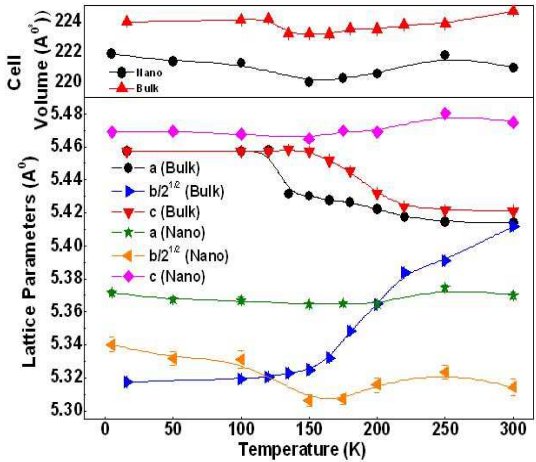}
\end{center}
\caption{Variation of lattice parameters and cell volume for bulk and nano LCMO (average particle size $\sim$ 15nm). Error bars, where not visible, are smaller than the symbols.}
\label{Fig5}
\end{figure}

\noindent
In Fig.~\ref{Fig5} we show the temperature evolution of the lattice parameters for the bulk as well as the nanoparticle sample with the smallest average particle size. In the same graph we also show the dependence of the unit cell volume. The variations of the lattice constants with temperatures are sharply different in the nanocrystals and the bulk sample. For the nanocrystals, throughout the temperature range studied, all the lattice constants  remain essentially  unchanged. There are no systematic changes in the lattice constants on cooling. At room temperatures, the nanocrystals have the a-axis  smaller  by $\approx$1\% compared to the bulk. The compaction in the b-axis is  $\approx$2\% while the c-axis expands by $\approx$1\%. At the lowest temperatures (5K) due to the large changes in the lattice constants of the bulk sample, the lattice constants a and c become comparable in the bulk samples while the b-axis becomes smaller. The absence of any temperature variation of the lattice constants of the nanocrystals show that the room temperature structure is indeed arrested in the nanocrystals. This arrest of the room temperature structure in the nanocrytals thus prevents the evolution of the low temperature charge ordered phase. 

\noindent
The lattice parameters of bulk LCMO (unlike those of the nanocrystals) display large changes in the region $T_{CO} < T < T_C$.  At 300K, the three axes have similar size. On cooling, the $b$ axis decreases drastically and the $a$ and $c$ axes increase correspondingly. The changes become more pronounced below 200K and they become nearly temperature independent for $T < T_{CO}$. These changes in the lattice parameters are associated with the structural changes which occur as the bulk sample undergoes the charge ordering transition and are characteristic of bulk LCMO. The observed changes in the lattice constants for the bulk sample match well with the data published on bulk samples before\cite{ref4}.

\begin{figure}[t]
\begin{center}\includegraphics[width=8cm,height=7cm]{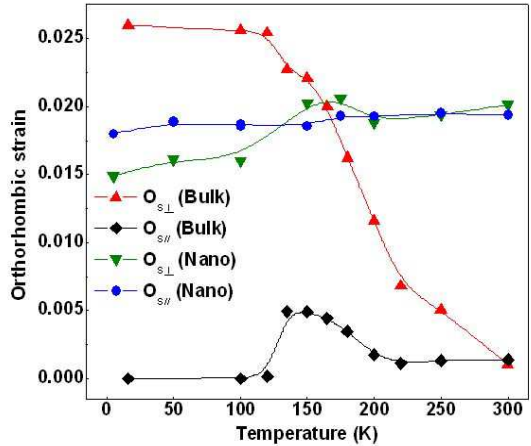}
\end{center}
\caption{Variation of the orthorhombic strains for bulk and nano LCMO(average particle size $\sim$ 15nm). Error bars, where not visible, are smaller than the symbols.}
\label{Fig6}
\end{figure}

\noindent
One important quantity that changes due to the size reduction are the orthorhombic strains ($O_{S_{\parallel}}$ and $O_{S_{\perp}}$)\cite{APL}. These are shown for the two samples in Fig.~\ref{Fig6}. The orthorhombic strain $O_{S_{\parallel}}$ gives the strain in the $ac$ plane and is defined as $O_{S_{\parallel}} = 2(c-a)/(c+a)$, while $O_{S_{\perp}}$ gives the strain along the $b$ axis with respect to the $ac$ plane and is defined as $O_{S_{\perp}} = 2(a+c-b\sqrt{2})/(a+c+b\sqrt{2})$. In the bulk sample, the largest change occurs in  $O_{S_{\perp}}$ which increases substantially as the sample is cooled and reaches a saturating value of $\approx$0.026 below the charge ordering temperature $T_{CO}$. On the other hand, $O_{S_{\parallel}}$ shows a rather small value over the whole temperature range while showing a small enhancement in the temperature range where the CO sets in but again becoming negligible  for $T < T_{CO}$. In the nanocrystal, since the lattice constants are more or less temperature independent, both $O_{S_{\parallel}}$ and $O_{S_{\perp}}$ are temperature independent over the whole temperature range and they have similar magnitudes. The orthorhombic strains that rapidly change in bulk samples at $T\approx T_{CO}$, is completely absent in the nanocrystals and as stated before the $O_{S_{\perp}}$ as well as  $O_{S_{\parallel}}$ remain locked at their room temperature values.  

\noindent
\subsection{MAGNETIC MEASUREMENTS}
\begin{figure}[t]
\begin{center}
\includegraphics[width=8cm,height=7cm]{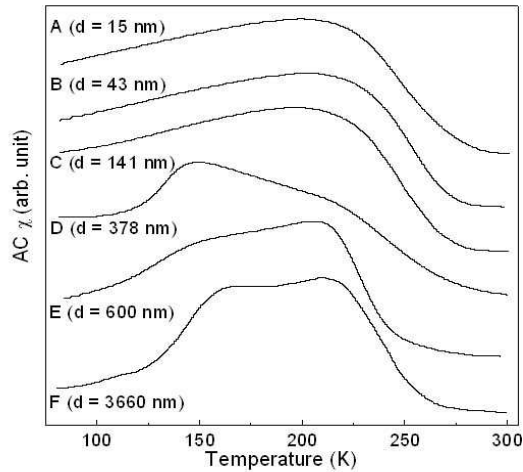}
\end{center}
\caption{A.C. suscpetibility vs. temperature for the LCMO samples having different particle sizes.}
\label{Fig7}
\end{figure}

\begin{figure}[t]
\begin{center}
\includegraphics[width=8cm,height=7cm]{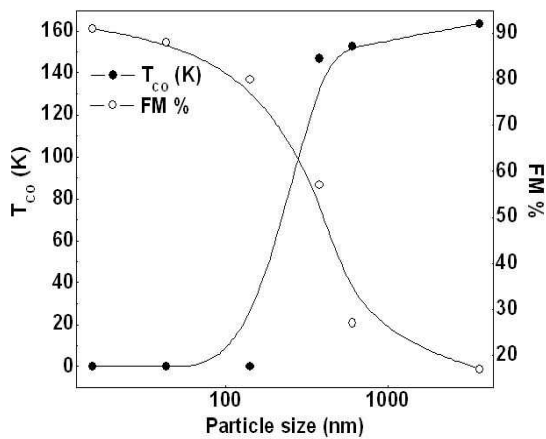}
\end{center}
\caption{Variation of $T_{CO}$ and FM $\%$ as a function of particle size.}
\label{Fig8}
\end{figure}

The structural changes that are observed in the nanocrystals are accompanied by distinct changes in the magnetic properties of the samples. Charge ordering in LCMO system is accompanied by spin ordering. In bulk LCMO, it has been reported that the spin ordering is an antiferromagnetic transition to CE type ordering with the Neel temperature $T_N \approx 155K$. This transition shows up as a drop in the susceptibility following the high temperature transition from paramagnetic to ferromagnetic state at $T_C \approx 225K$. Fig.~\ref{Fig7} shows the temperature dependence of the low field a.c. susceptibilities (taken by  mutual inductance bridge) of LCMO samples of different sizes ranging from the highest (3660 nm) to the lowest (15 nm) particle sizes. All the samples undergo a transition from paramagnetic to ferromagnetic state at around 230K-250K where the $T_C$ is identified by the inflection points in the dM/dT versus T plots. Interestingly, the $T_C$ determined by the low field a.c. suscpetibility increases on size reduction signifying strengthening of the ferromagnetic interaction on size reduction. As the temperature is decreased further, the susceptibility of the bulk sample (average particle size $\approx 3660nm$) starts to decrease at $T \approx 150K$ and the transition is mostly over by $T \approx 135K$. This is the signature of the onset of the Antiferromagentic (AFM) order accompanying the CO, as has been seen earlier in bulk samples\cite{Mathur}. Such a drop in the suscpetibility at lower temperatures are absent in nanocrystals of size $\leq$ 150nm. This would imply that the AFM order that accompanies the CO is absent in nanocrystals with sizes smaller than 150nm. In Fig.~\ref{Fig8} we plot $T_{CO}$ and the ferromagnetic fraction as determined by the magnetization (magnetization shown later on) as a function of particle size. It can be seen clearly that both $T_{CO}$ as well as the FM fraction show a sharp transition when the particle size goes below 150nm. The size reduction thus inhibits formation of the CO ground state and the ferromagnetic state is stabilized.

\begin{figure}[t]
\begin{center}
\includegraphics[width=8cm,height=7cm]{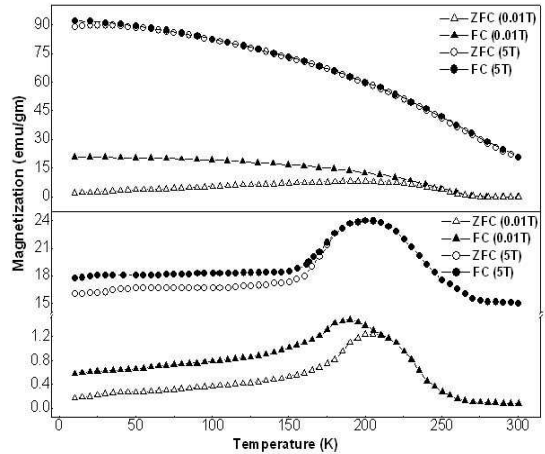}
\end{center}
\caption{Magnetization vs. temperature for the nanoparticle sample (upper panel) and bulk sample (lower panel) of LCMO under a magnetizing field of 0.01T and 5T.}
\label{Fig9}
\end{figure}

\noindent
The magentization data supports the observation as made in the a.c. susceptibility. The magnetization data taken on the "bulk" sample as well as on the nanocrystals are shown in Fig.~\ref{Fig9} at two fields (0.01T and 5T). The bulk sample (lower panel) shows the transition from paramagnetic to ferromagnetic state where the magnetization rises, and then on further cooling as the AFM order sets in, it falls as seen in the a.c. susceptibility data. At lower field $(\approx 0.01T)$ there is a clear separation of the ZFC and FC data which occurs close to the temperature where the magnetization shows a peak. This is a signature of irreversibility that may arise due to remanent spin disorder in the sample. The magnetic moment at 10K at 5T field is $\approx 0.6\mu_B/f.u.$. This is substantially less than that expected from fully ferromagnetically aligned moments. The AFM order that sets in along with the CO transition is thus stable down to the lowest temperature in the bulk sample in a field of 5T. In contrast to the bulk sample the nanocrystals do not show any magnetic transition below the transition from paramagnetic to ferromagnetic state. The magnetization of the nanocrystals at H=5T rises monotonically (for both FC and ZFC cases) and approaches towards saturation at the lowest most temperature. The magnetization at 10 K is $\sim$ 3.18$\mu_{B}/f.u.$ This is $\sim$ 91{\%} of $M_{S}$ (=3.5$\mu_{B}/f.u.$, calculated assuming full ferromagnetic alignment of the spins  for $\mathrm{La_{0.5}Ca_{0.5}MnO_3}$). For the intermediate samples data are not shown to avoid repetition. The saturation moment determined from the magnetization has been shown in Fig.~\ref{Fig8} before.

\noindent
A plot of inverse susceptibility (1/$\chi$) vs temperature for the  samples in the paramagnetic region show a  Curie Weiss law ($\chi = C/(T - \theta$)), where $C$ is the Curie constant and $\theta$ is the Weiss temperature. The Curie temperatures in all the  samples are ferromagnetic. From the Curie constant $C$, we could calculate the effective magnetic moment $\mu_{eff}$. For the 3660nm bulk sample $\mu_{eff}\approx 4.14\mu_{B}$ and for the 15nm nanocrystal  $\mu_{eff}\approx 4.36\mu_{B}$. This establishes that the paramagnetic state of LCMO is not  much affected by the particle size. The main effect arises on cooling below the first transition from paramagnetic to ferromagnetic state and the subsequent presence or absence of the antiferromagnetic transition. The magnetic moment enhancement at low temperatures in the nanocrystals distinguishes the two ground states seen in the bulk sample and the nanocrystals. 

\begin{figure}[t]
\begin{center}
\includegraphics[width=8cm,height=7cm]{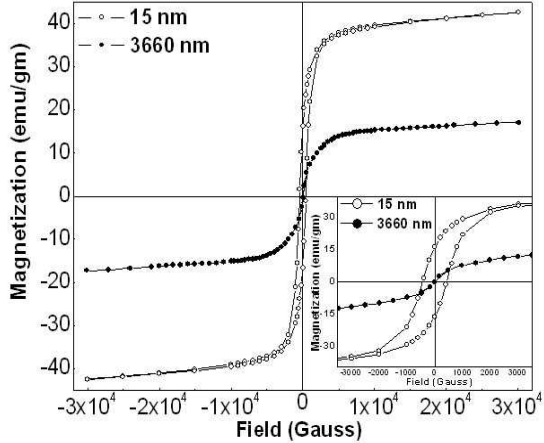}
\end{center}
\caption{Magnetization vs. field for bulk and nano LCMO at a temperature of 5K. The inset shows an expanded portion of the MH loop near the origin.}
\label{Fig10}
\end{figure}

\noindent
In Fig.~\ref{Fig10} we show the M vs. H curves for the bulk (average particle size $\sim$ 3660nm) and nano particle (average particle size $\sim$ 15nm) samples taken at 5K. Compared to the bulk sample, which shows no hysteresis in the MH loop, the nano particle sample shows a considerable hysteresis with a coercive field of $\sim$ 0.05T and a remanence magnetization of $\sim$ 16emu/gm (see inset of Fig.~\ref{Fig10}). In contrast there is no remanence or coercive field in the larger size particles (bulk sample).

\noindent
The magnetization data show that the nanocrystals retain their ferromagnetic state as they are cooled to lower temperatures unlike the bulk sample that shows the antiferromagnetic transition. The data also show that there is a significant enhancement of the ferromagnetic moment in the nanocrystals. 

\noindent
\subsection{ ELECTRICAL TRANSPORT AND MAGNETORESISTANCE}
Electrical resistivity provides strong evidence for the charge and orbital ordering transition when the sample enters an insulating state on charge ordering. Often the charge ordering transition is reflected as a change in slope in the resistivity plot. CO state also shows a very strong magnetoresistance in some of the manganites as the insulating state is destabilized by an applied field which leads to creation of ferromagnetic metallic state. In the present investigation, particularly for the nanocrystals, the particle size being small the transport experiments have interference from the presence of a large contribution of grain boundaries in the electrical as well as magnetotransport behavior. However, we show below that in spite of the interference one can distinguish the electrical and the magnetotransport behaviors in samples that have large grains ($\sim 3660nm$) and those which have nanocrystals ($\approx 15nm$). 

\begin{figure}[t]
\begin{center}
\includegraphics[width=8.2cm,height=7.5cm]{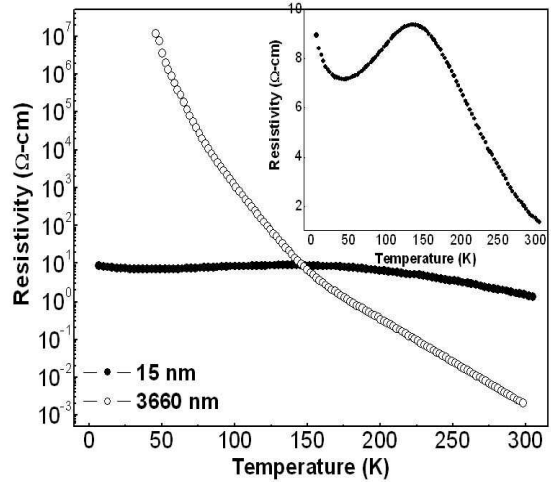}
\end{center}
\caption{Resistivity vs. temperature for LCMO (15nm) and bulk LCMO. Inset shows the resistivity of the nano particle sample on an extended scale.}
\label{Fig11}
\end{figure}

\begin{figure}[t]
\begin{center}
\includegraphics[width=8.2cm,height=7.5cm]{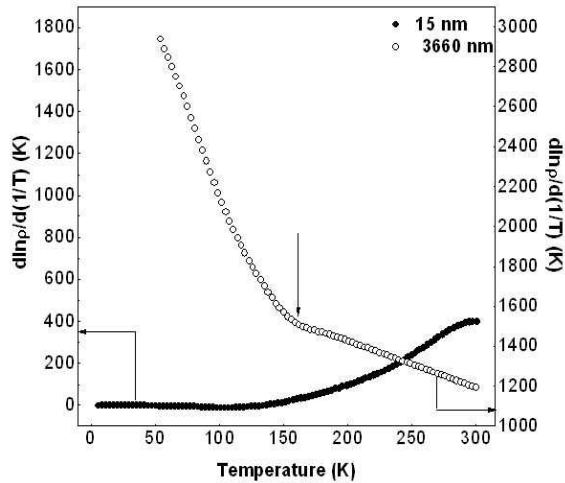}
\end{center}
\caption{$\frac{dln\rho}{d(1/T)}$ vs. temperature for LCMO (15nm)  and bulk LCMO.}
\label{Fig12}
\end{figure}

\noindent
In Fig.~\ref{Fig11} we show the behavior of the resistivities of the two samples (3660nm(bulk)  and 15nm samples) as a function of temperature from 4.2K to 300K. The data are shown in the log scale. In nanoparticles,  the resistivity at room temperature is about 3 orders higher than that of the bulk sample. In this temperature range both the samples are in charge and orbitally disordered paramagnetic state. The higher resistance of the sample with nanocrystals reflects the enhanced grain boundary contribution. In the bulk sample, however, the resistivity rises rapidly as the sample is cooled through the CO transition and at $50K$ the resistivity in the bulk sample becomes more than 6 orders higher than that in the nanoparticles. The bulk sample shows an insulating behavior throughout the temperature range. The charge ordering transition is reflected as a change in slope in the resistivity plot in the bulk sample. This can be seen in Fig.~\ref{Fig12} where we plot $dln\rho/d(1/T)$ which gives the transport activation energy at temperature T. For the bulk sample we find a clear change in slope (at T $\approx 150K$) showing hardening of the transport gap on cooling through the CO temperature. 

\noindent 
The transport data of the sample  with 15nm nanocrystals are also shown in  Fig.~\ref{Fig11}. In contrast to the sample with 3660nm particle size, these samples show much less sensitivity to temperature and the resistivity does not vary much on cooling (within one order). The resistivity of this sample shows a shallow peak at $T_{P}\sim$ 160 K (seen more clearly in the inset of Fig.~\ref{Fig11}). In conventional ferromagnetic manganite $\mathrm{La_{1-x}Ca_{x}MnO_3}$ with $x \approx 0.3$,  a metal-insulator transition occurs close to the ferromagnetic transition where the material shows transition from a polaronic insulating paramagnetic state to a ferromagnetic metallic state. Thus a peak in the resistivity occurs close to $T_C$. For this sample the magnetic data shows a $T_{C}\sim 225K$. Thus $T_P$ is significantly lower than $T_C$. Such a behavior is expected when the material is not a homogeneous metallic phase but it is percolative in nature because of a large number of grain boundaries or presence of an insulating phase co-existing with the metallic phase.
The resistance again starts to rise  below 55 K, presumably due to predominant contribution of the insulating grain boundaries to the overall current transport. Such behaviour has been seen in nanostructured ferromagnetic films of manganites \cite{ref24}. The plot of $\frac{dln\rho}{d(1/T)}$ for the nano particle sample (Fig.~\ref{Fig12}) shows that unlike the bulk sample, there is no clear change at around 150K, which one would expect if the system showed a CO transition. If anything, it shows a gradual decrease in slope. An analysis of the resistivity data shows that there is indeed a signature of suppression of the CO insulating phase in the nanoparticles and the phase that forms  is   metallic in nature although a percolative type of bad metal  because of the existence of a large number of grain boundaries or even a co-existing insulating phase. 

\begin{figure}[t]
\begin{center}
\includegraphics[width=8.2cm,height=7.5cm]{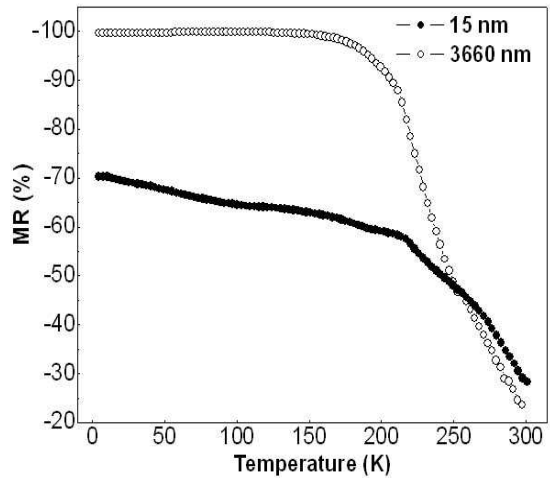}
\end{center}
\caption{MR as a function of temperature for LCMO (15nm) and bulk LCMO}
\label{Fig13}
\end{figure}

\indent
The MR [defined as $100(\rho(H) - \rho(0))/\rho(0)$ (with H = 10T)] for both the bulk (average particle size $\sim$ 3660nm) and nano particle sample (average particle size $\sim$ 15nm) are plotted as a function of temperature in Fig.~\ref{Fig13}. In both the samples the MR is negative although the value as well the temperature dependence are qualitatively different. The insulating state of the CO bulk sample is completely suppressed on application of 10T magnetic field. This is due to magnetic field induced destabilization of the CO state that has been seen in many CO systems. The MR at the lowest temperature is nearly 100{\%} for the bulk sample for $T < 100K$.  In contrast, in the nanocrystal sample the MR increases slowly as it is cooled and reaches the limiting value of   $\sim$ 70{\%} . The behavior of the MR in the nanocrystal is similar to that seen in nanostructured films of ferromagnetic manganites\cite{ref25,ref26} which arises predominantly due to the grain boundary contributions. The MR data of the two samples also corroborate the earlier observation about the predominant phases present in  the two  samples.

\noindent
\section{DISCUSSIONS}
In this paper, we have studied how the crystal structure and the physical properties of $\mathrm{La_{0.5}Ca_{0.5}MnO_3}$ evolve on reducing the particle size by more than 2 orders of magnitude. We find a drastic change in the structure of the system on size reduction. The magnetic as well as magneto-transport behavior of the nanocrystals also show qualitative changes in the nanocrystals. In the nanocrystals the CO state does not develop and the ferromagnetic order is stable down to lowest temperatures. At room temperature the nanocrystals are more distorted than the bulk sample, as is evidenced from the higher values of the orthorhombic strains at room temperature in the nanoparticle sample. However, this distortion is temperature insensitive as is clear from Fig.~\ref{Fig5} and Fig.~\ref{Fig6}. These observations bring forward some questions. First, why the crystal structure of the nanoparticles is different and the room temperature structure does not evolve with temperature unlike the bulk crystal. Second, what prevents the CO to develop on cooling and whether the structure has a role to play.

\indent
To explain this change in the structure on size reduction we make the proposition that the surface pressure makes the nanocrystals behave as material under high hydrostatic pressure which not only changes the structure, but also "locks" the room temperature structure. Below, we justify this proposition. If we assume our particles to be spherical in shape, then the surface pressure acting on the particles is given by $P_{s} = 2S/d$, where $d$ is the diameter of the particle and $S$ is the surface tension. The exact value of $S$ for manganites is not known, but for perovskite oxide titanates, $S \approx 50N/m$\cite{ST}. Putting in the values, we find that a pressure of $P_{s} \approx 6 GPa$ acts on the sample with average particle size $\sim$ 15nm. It is clear that for particles of larger size ($\geq 100$nm) the surface pressure will be very small ($\leq 1 GPa$) and thus will be of no consequence. The hydrostatic pressure will lead to reduction in cell volume as has been observed. From the observed cell volume change ($\sim 1.7\%$) using the typical  bulk modulus of manganites as $\sim 150GPa$ we find that a hydrostatic pressure of $\sim 6GPa$ will lead to a  reduction in cell volume of $\sim 4\%$ which is similar to but larger than the observed reduction. This simple argument explains that the surface effect can indeed produce enough hydrostatic pressure to explain the change in the unit cell volume. Since the effect of surface pressure is to produce an effective hydrostatic pressure it will be worthwhile to make connection with recent direct investigation of crystal structure under applied hydrostatic pressure as reported recently \cite{HP}.  Our data for the nanoparticle sample (which we consider to be under an effective pressure of $\sim$ 6$GPa$) matches very well with the crystal structure data of bulk sample measured under directly applied hydrostatic pressure of the same magnitude. Thus comparison to the hydrostatic pressure data establishes that the  nanocrystals samples are under an effective hydrostatic pressure created by surface pressure due to its small size. It is this effective pressure that causes the crystal structure to deviate from the bulk structure.

\noindent
The related issue is what causes the room temperature crystal structure of the nanocrystals not to evolve with temperature as the sample is cooled unlike the bulk sample. We suggest that the effective hydrostatic pressure created by the size the reduction acts like a "clamp" that frrezes-in the room temperature structure. The structural evolution on cooling in the bulk sample involves expansion of $a$ and $c$ axes and contraction of $b$ axes. It is likely that the hydrostatic pressure prevents the lattice expansion. As a rough estimate using the $\sim 1.7\%$ unit cell compaction as a strain due to the effective hydrostatic pressure of $6 GPa$ we find that the energy involved is $\approx 60$meV which is larger than the thermal energy. This will justify why the effective hydrostatic pressure $P_{s}$ will freeze-in the room temperature structure.

\noindent
The effect of size reduction is to cause a change in the lattice structure. The onset of CO needs a particular type of crystal structure (or distortion) to support it. It is thus tempting to connect the absence of CO in nanocrystals to the structural factors. One reason can be that the particular type of orthorhombic distortion where $O_{S_{\perp}}$ is substantially larger than $O_{S_{\parallel}}$ is needed for the CO to set in and this is absent in the nanocrystals where 
$O_{S_{\perp}}\approx O_{S_{\parallel}}$. This may prevent the CO to set in. Also, the development of CO needs creation of a modulated structure and a supercell as has been seen in bulk samples of $\mathrm{La_{0.5}Ca_{0.5}MnO_3}$\cite{ref4}. The propagation vector of the CO modulated structure is (1/2+$\epsilon$,0,0), where $\epsilon\approx 0.01$. This implies that the periodicity of the supercell is $\approx$ 200$a\approx$106nm. Thus, if the particle size is less than $\sim$100nm the supercell modulation needed for the CO cannot develop. In our case the size of the nanocrystals is more than 7 times less than this value. It appears that the  absence of a supercell modulation can be another cause why the CO does not  set-in in the nanocrystals.

\noindent
It will be worthwhile to explore the other possible mechanisms which might lead to the suppression of the charge ordered state, and investigate whether they have any relevance in our case. One of the likely mechanisms can be site or surface disorder. While disorder can most definitely lead to a destabilization of the charge ordered state, it is not clear why this should lead to an enhancement and strengthening of the ferromagnetic interaction as we observe in the nanocrystals. In fact, both site as well as surface disorder should lead to a decrease in the ferromagnetic $T_C$ and the ferromagnetic moment\cite{site,surface}. In our samples, we find just the opposite trend i.e. an enhancement in the ferromagnetic $T_C$ as well as an increase in the ferromagnetic moment as the particle size is reduced. In fact, we note here that the nanocrystals do not have much spin disorder. The high field differential susceptibility ($\partial M/\partial H$) which can be taken as a measure of the spin disorder is the same in both the nanoparticles and the bulk samples ($\partial M/\partial H \approx 0.025$emu/gm.kGauss at 3T). While disorder as seen in the grain boundary or grain surfaces appear to be an unlikely cause one cannot rule out random local strain as arising from inhomogeneous strain as a factor. This can act as a random field and that can indeed destabilize the CO state. At present we cannot rule this out and it may happen that the random field can act in tandem with the surface pressure arising from size reduction and lead to destabilization of the CO phase and creation of the FM order. It may be emphasized that the X-ray data show that the change occurs in the "bulk" of the nanocrystals and not in the surface.

\noindent
At the end it may be pointed out that manganites in nanoscale are somewhat novel because it may be one of the few known systems where the metallic state (with ferromagnetic interaction) is stabilized by size reduction. In almost all the reported oxide systems the size reduction destabilizes the metallic state and one obtains transition to an insulating state\cite{LSMOtransport}. Manganites have competing interactions with almost equal strengths. The size tuning thus provides a subtle change in the balance between relative strengths leading to destabilization of one phase. This investigation establishes that the ground state property of manganites can be tuned by size reduction and also suggests that the tuning may actually occur due to change in structural parameters.

\noindent
\section{CONCLUSIONS}
In  summary,  we  have  studied, in details, the  effect  of  size  reduction  on the crystal structure and physical properties of   $\mathrm{La_{0.5}Ca_{0.5}MnO_3}$ nanocrystals  which were synthesized  using  the  polymeric  precursor route. The size was reduced down to an average size of $15$nm by using chemical methods. It was observed that the size reduction leads to change in crystal structure and the room temperature structure is arrested so that structure does not evolve on cooling unlike bulk samples. The change in the structure was ascribed to an effective hydrostatic pressure created by surface pressure which not only changes the structures but causes the room temperature to freeze in. The size reduction does not allow the long supercell modulation needed for the CO to set-in. The magnetic and transport measurements also show that the CO does not occur when the size is reduced below a critical size. The nanoparticle samples show enhanced ferromagnetic moment and metallic type conductivity. Our investigation establishes a structural basis for the destabilization of CO state in nanocrystals. Though the experiment has been carried out in the specific context of $\mathrm{La_{0.5}Ca_{0.5}MnO_3}$, it is not unreasonable to expect similar behavior in other half-doped manganites. In fact the concept, that the surface pressure can create an effective hydrostatic pressure and that can act as a change agent, may be applicable in other systems whose properties can change substantially with moderate hydrostatic pressure.

\noindent
\section{ACKNOWLEDGMENTS}
Discussions on structure of perovskite oxides with Prof. D. Pandey, Banaras Hindu University, India is gratefully acknowledged. The authors thank the Unit of Nanoscience, Indian Association for the Cultivation of Science, Kolkata, India for TEM support. The MR measurements were done at the DST Facility for Low Temperature and High Magnetic Field at IISc, Bangalore, India. We thank Dr.Hermann Emerich of the E.S.R.F, Grenoble, France for help during data acquisition and Dr.P.Chowdhury, CGCRI, Kolkata, India for his help in the Rietveld refinements of the data. We also thank the Department of Science and Technology, Govt. of India for financial support in the form of a Unit and sponsored scheme. TS also thanks UGC, Govt. of India for fellowship. BG acknowledges financial support from the Department of Science and Technology, Govt. of India under Women Scientist Scheme.

\end{document}